\documentclass[11pt]{article}
\usepackage{a4wide}
\usepackage{cite}

\usepackage{amssymb}
\usepackage{amsmath}
\usepackage{epsfig}
\usepackage{xspace}

\newcommand{\LettTitle}{%
The gluon splitting function at moderately small $\boldsymbol{x}$
}

\newcommand{\LettAbstract}{
  It is widely believed that at small $x$, the BFKL resummed gluon
  splitting function should grow as a power of $1/x$. But in several
  recent calculations it has been found to \emph{decrease} for
  moderately small-$x$ before eventually rising.  We show that this
  `dip' structure is a rigorous feature of the $P_{gg}$ splitting
  function for sufficiently small $\as$, the minimum occurring
  formally at $\ln 1/x \sim 1/\sqrt{\as}$.  We calculate the
  properties of the dip, including corrections of relative order
  $\sqrt{\as}$, and discuss how this expansion in powers of
  $\sqrt{\as}$, which is poorly convergent, can be qualitatively
  matched to the fully resummed result of a recent calculation, for
  realistic values of $\as$.  Finally, we note that the dip position,
  as a function of $\as$, provides a lower bound in $x$ below which
  the NNLO fixed-order expansion of the splitting function breaks down
  and the resummation of small-$x$ terms is mandatory. }

\newcommand{\LLx}{LL$x$\xspace}
\newcommand{\NLLx}{NLL$x$\xspace}

\newcommand{\NLLB}{NLL$_\mathrm{B}$\xspace}
\newcommand{\as}{\alpha_s}              
\newcommand{\asb}{\bar{\alpha}_s}       
\renewcommand{\ln}{\log}
\newcommand{\order}[1]{\mathcal{O}\left(#1\right)}
\newcommand{\eff}{{\mathrm{eff}}}
\newcommand{\om}{\omega}
\newcommand{\omc}{\omega_c}
\newcommand{\oms}{\omega_s}
\newcommand{\nf}{n_f}
\newcommand{\NC}{N_c}
\newcommand{\MSbar}{\overline{\mbox{\scriptsize MS}}}

\begin{document}

\titlepage
\begin{flushright}
DESY 03--185 \\ DFF 409/11/03\\  LPTHE--03--34 \\ hep-ph/0311325 \\
November 2003
\end{flushright}

\vspace*{1.0in}
\begin{center}
{\Large \bf
\LettTitle
}\\
\vspace*{0.4in}
M.~Ciafaloni$^{(a)}$,
D.~Colferai$^{(a)}$,
G.P.~Salam$^{(b)}$
and A.M.~Sta\'sto$^{(c)}$ \\
{\small
\vspace*{0.5cm}
$^{(a)}$ {\it  Dipartimento di Fisica, Universit\`a di Firenze,
 50019 Sesto Fiorentino (FI), Italy}; \\
\vskip 2mm
{\it  INFN Sezione di Firenze,  50019 Sesto Fiorentino (FI), Italy}\\
\vskip 2mm
$^{(b)}$ {\it LPTHE, Universities of Paris VI \& VII and CNRS,
75252 Paris 75005, France}\\
\vskip 2mm
$^{(c)}$ {\it Theory Division, DESY, D22603 Hamburg};\\
\vskip 2mm
{\it H.~Niewodnicza\'nski Institute of Nuclear Physics, Krak\'ow, Poland}\\
\vskip 2mm}
\end{center}

\vskip1cm
\begin{abstract}
  \LettAbstract
\end{abstract}
\newpage

\section{Introduction}

A major effort is currently under way to push the precision of
DGLAP~\cite{DGLAP,NLODGLAP1,NLODGLAP2} splitting functions to
next-next-to-leading order (NNLO) accuracy~\cite{NS3loop}. One of the
main applications of such an effort could be to improve the
description of the small-$x$ parton distributions, which with the
current NLO evolution suffer from pathologies such as negative gluon
distributions and predictions of a negative
$F_L$~\cite{MRSTNLO,CTEQNLO}. Furthermore a good knowledge of
small-$x$ parton distributions will be ever-more relevant as collider
energies are increased, for example at the LHC or a possible VLHC,
which will be able to probe small-$x$ kinematic regions unexplored
even at HERA. 

However a question that remains to be understood is that of the domain in
which fixed order expansions are sufficiently convergent as to be
reliable.  Indeed it is known that at small $x$, there are large
logarithmic enhancements of the splitting function at all
orders~\cite{BFKL,JKCOL}, leading formally to the breakdown of the 
convergence of the series for $\as \ln 1/x \sim1$ and it has been
argued~\cite{MRSTNNLOUncert} that there is evidence in the
data~\cite{HERA} for the presence of some such terms.

Much effort has been devoted in recent years to resumming these
logarithmically enhanced terms, which are expected to lead to a rise
at small $x$, as a power of $x$, for the gluon-gluon splitting
function, $xP_{gg}(x)$.  It turns out however that the \LLx summation,
$\as^n \ln^{n-1}1/x$ rises much too steeply~\cite{EHW,BF95} to be
compatible with the more gentle rise of the $F_2$ data~\cite{HERA}.
On the other hand, the inclusion of the \NLLx terms
$\as^n\ln^{n-2}1/x$ --- extracted from the \NLLx kernel
eigenvalue~\cite{NLLFL,NLLCC} and based on several Regge-gluon
vertices~\cite{RGvert} and on the $q \bar{q}$ cluster
~\cite{QQvertCC,QQvertFFFK} --- leads at moderately small $x$ to a
negative splitting function~\cite{BLUMVOGT,ROSS98}.  Since that
discovery, there has been investigation of the origin of these
problems, and various approaches have been proposed to estimate yet
higher
orders~\cite{Salam1998,CC,CCS1,CCSSkernel,SCHMIDT,FRSV,THORNE,ABF2000,ABF2001,ABF2003,ABFcomparison},
the most successful of them being based on a simultaneous treatment of
small-$x$ and collinear logarithms.

A surprising observation, common to all these approaches, is that in
the phenomenologically relevant, moderately small-$x$ region, the
splitting function actually \emph{decreases}, while the power-like
rise is delayed to somewhat smaller values of $x$ (resummed curve of
figure~\ref{fig:pgg}~\cite{CCSSkernel}, which has been found to be
rather close to a splitting function fitted to the $F_2$ data
\cite{ABF2000,ABFcomparison}). The question arises therefore of
whether the resulting `dip' structure is a well-defined property of
the gluon splitting function, or instead perhaps an artefact of the
particular schemes used to `improve' the small-$x$ hierarchy.  The
purpose of this letter is to show that the dip has a simple origin,
specifically in the structure of the first few terms of the
perturbative series, possibly matched to a resummed behaviour at
smaller $x$ values.

More precisely (Sec.~\ref{sec:loworders}), in the formal
limit of small $\as$, the dip is a consequence of an interplay between
different fixed orders, and one finds that the simple fixed-order
hierarchy breaks down not for $\as \ln 1/x \sim 1$ as widely expected,
but rather for $\as \ln^2 1/x \sim1$. The result is that the
properties of the dip can be described in terms of a series in powers
of $\sqrt{\as}$.  For phenomenologically relevant values of $\as$
though, this series in $\sqrt{\as}$ turns out to be very poorly
convergent.  Instead we find that quite simple resummation arguments,
presented in section~\ref{sec:resummation}, still enable us to gain
some quantitative understanding of the dip properties.

\section{Low perturbative orders and $\boldsymbol{\sqrt{\as}}$-expansion}
\label{sec:loworders}

Let us start by recalling the structure of the \LLx terms of the
$xP_{gg}(x)$ splitting function,
\begin{equation}
   A_{n,n-1} \,\asb^{n}\, \ln^{n-1} \frac1x\,,\qquad\quad (n\ge 1)\,,
\end{equation}
where $\asb = \as \NC/\pi$. A number of the lower order terms in the
series are absent, $A_{21} = A_{32} = A_{54} = 0$, while
\begin{equation}
  \label{eq:Acoeffs}
  A_{10} = 1\,,\qquad A_{43} = \frac{\zeta(3)}{3}\,,\qquad A_{65}
  =\frac{\zeta(5)}{60}\,,\qquad
  \ldots
\end{equation}
Since these and all further terms are positive, the \LLx splitting
function grows monotonically as $x$ decreases.  The \NLLx terms can be
written as
\begin{equation}
   A_{n,n-2} \,\asb^{n}\, \ln^{n-2} \frac1x\,,\qquad\quad (n\ge 2)\,,
\end{equation}
where the first few coefficients are~\cite{NLLFL,NLLCC}
\begin{subequations}
  \label{eq:Bcoeffs}
\begin{align}
  A_{20} &= -\frac{\nf}{6\NC}\left(\frac53 + \frac{13}{6\NC^2}\right)
  \,,\\
  A_{31} &= -\frac{395}{108} + \frac{\zeta(3)}{2} + \frac{11\pi^2}{72} -
  \frac{\nf}{4\NC^3} \left(\frac{71}{27} - \frac{\pi^2}{9}\right)
  \simeq -1.548 - 0.014\nf
  \,,\\
  A_{42} &= -4.054 - 6.010\,b - 0.030 \nf = -9.563 + 0.303 \nf\,, \qquad \ldots
\end{align}
\end{subequations}
and $b = \frac{11}{12} - \frac{\nf}{6\NC}$ is the first beta-function
coefficient.  These coefficients are given in the $Q_0$
scheme~\cite{Q0} and for renormalisation scale $\mu=Q$.
They come from a simple expansion of the \NLLx kernel eigenvalue, and ---
notably $A_{31}$ and $A_{42}$ --- can be traced back to early calculations of
\NLLx gluon vertices~\cite{RGvert} and of the $q \bar{q}$
cluster~\cite{QQvertCC,QQvertFFFK}. They include --- in particular $A_{42}$ ---
the running coupling effects, which are part of the \NLLx corrections. In the
$\MSbar$ scheme only the $\nf$ parts of $A_{20}$ and $A_{31}$ will differ, while
from $A_{42}$ onward the $\nf$ independent part will differ as well. Because of
the zeroes in the LL coefficients, $A_{31}$ and $A_{42}$ are independent of the
choice of $\mu$.

\begin{figure}[tb]
  \centering
  \includegraphics[width=0.6\textwidth]{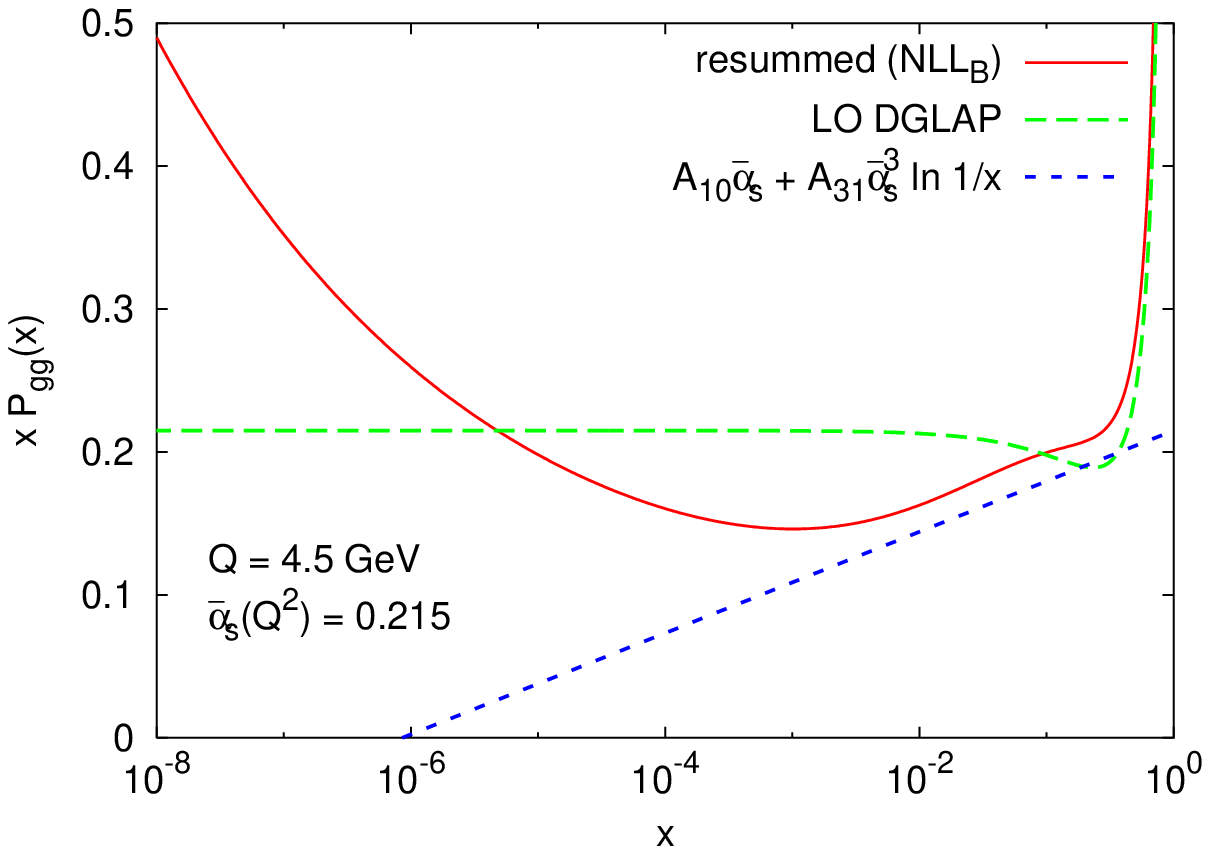}
  \caption{The $x P_{gg}(x)$ splitting function. The resummed (\NLLB)
    curve corresponds to scheme B of~\cite{CCSSkernel}.}
  \label{fig:pgg}
\end{figure}

The resummation hierarchy as written above in terms of \LLx and \NLLx
terms is intended to be applied when $\as \ln 1/x$ is of order $1$,
while $\as \ll 1$ and $\ln1/x \gg1$. Let us however examine an
intermediate small-$x$ limit in which $\ln 1/x \gg 1$ but $\as \ln 1/x
\ll 1$ (the precise region will be better specified shortly).

Because the \LLx coefficients $A_{21}$ and $A_{32}$ are zero, the
lowest order term with $\ln 1/x$ enhancement is the \NLLx term $A_{31}
\as^3 \ln 1/x$, which is NNLO in the usual DGLAP perturbative
expansion. Since $A_{31}$ is negative it will lead to an initial
\emph{decrease} of the splitting function and at some sufficiently
small value of $x$ the NNLO gluon splitting function~\cite{vNV} will
become negative, as shown in figure~\ref{fig:pgg}, where we have
included the small-$x$ part of the NNLO $xP_{gg}(x)$, $A_{10}\asb +
A_{31} \asb^3 \ln1/x$ ($A_{20}=0$ in the particular scheme used in the
figure~\cite{CCSSkernel}).

At N$^3$LO, order $\as^4$, both \LLx and \NLLx terms are present.
Since we are in the regime of $\ln 1/x \gg1$, the \LLx $\as^4 \ln^3
1/x$ term will clearly dominate over the \NLLx $\as^4 \ln^2 1/x$
term. What is interesting however is the interplay between the
negative \NLLx $\as^3 \ln 1/x$ term and the positive \LLx $\as^4 \ln^3
1/x$:
\begin{equation}
  \label{eq:firstterms}
  x P_{gg}(x) = \mathrm{const.} + A_{31}\, \asb^3 \ln
  \frac1x + A_{43}\, \asb^4 \ln^3\frac1x + \cdots\,,
\end{equation}
where the constant term includes $A_{10}\asb$ and $A_{20}\asb^2$
contributions, and at each order in $\as$ we have written only the
term with the strongest $\ln 1/x$ dependence. Since $A_{43}$ is
positive and has stronger $\ln x$ dependence than the negative
$A_{31}$ term, the splitting function as written in
\eqref{eq:firstterms} will eventually start rising. The $A_{31}$ and
$A_{43}$ terms will be of the same order when $\as \ln^2 1/x \sim 1$,
and  the splitting function of eq.~\eqref{eq:firstterms}
will have a minimum at 
\begin{equation}
  \label{eq:min0}
  \ln \frac1{x_{\min}} = \sqrt{-\frac{A_{31}}{3A_{43}}\frac1{\asb}}\,.
\end{equation}
The appearance of this minimum for $\as \ln^2 1/x \sim 1$ suggests
that it may be of use to examine an alternative classification of the
series, in which we consider all terms that are of similar magnitude
when $\as \ln^21/x$ is of order one,\footnote{This is a
  double-logarithmic classification, however one should bear in mind
  that the perturbative series itself contains at most single
  logarithms --- our study of powers of $\as \ln^21/x$ therefore just
  represents a particular way of reclassifying terms in the
  single-logarithmic perturbative expansion.}
\begin{equation}
  \label{eq:DLclass}
  A_{k,2k-5}\, \asb^k \ln^{2k-5} \frac1x\,,\qquad\quad (3\le k\le 4)\,.
\end{equation}
One finds that there are only terms with $k=3,4$, since lower values
of $k$ would be associated with negative powers of $\ln 1/x$, while
higher values of $k$ would be super-leading in the usual \LLx
classification.  In other words the two terms, $\as^3 \ln 1/x$ and
$\as^4 \ln^3 1/x$, that we have examined so far provide the full
leading contribution for $\as \ln^21/x \sim1$.  

\begin{figure}[htbp]
  \centering
  \includegraphics[width=0.6\textwidth]{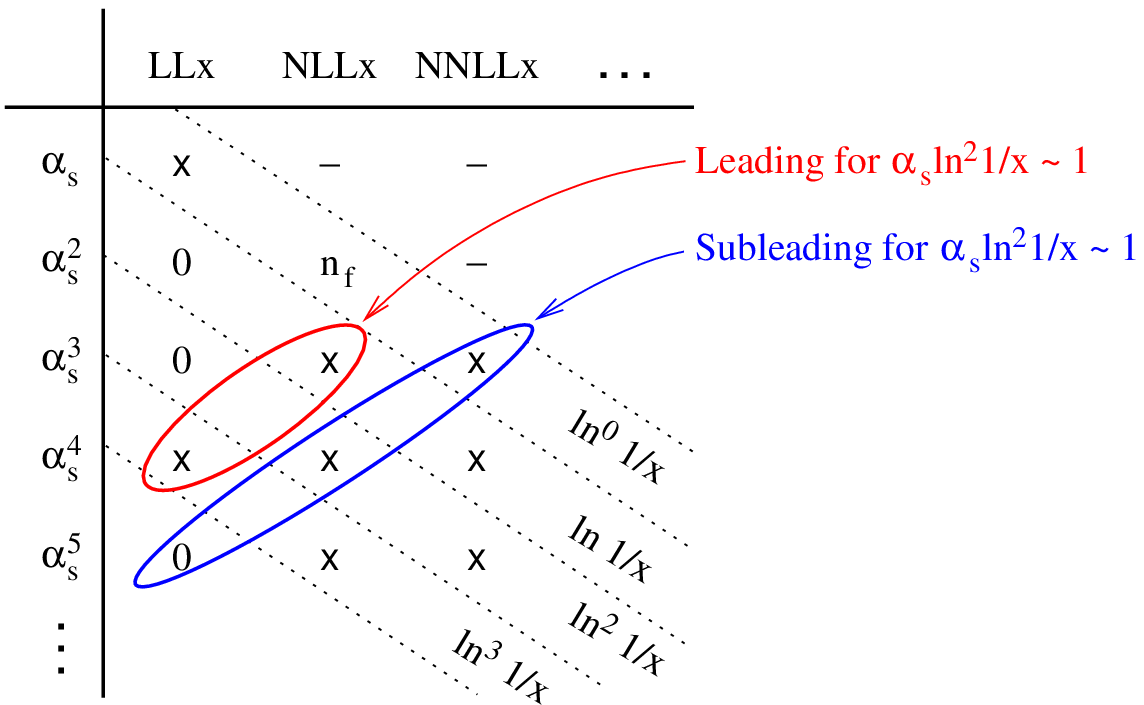}
  \caption{Representation of different classifications of
    logarithmically enhanced terms. Symbols `\textsf{x}' indicate
    terms that are present; `$0$' indicates terms that could have been
    present but are zero; `$\nf$' indicates a term whose only
    non-vanishing part is proportional to $\nf$; a dash indicates
    terms which do not exist by definition.}
  \label{fig:table}
\end{figure}

This is illustrated in figure~\ref{fig:table}, which shows various
possible classifications of logarithmically enhanced terms. Rows
correspond to a given power of $\as$; columns to a given
single-logarithmic order (\LLx, \NLLx, and so on); terms on a same
downward going diagonal line (reading from left to right) all have
the same power of $\ln x$.

Terms on upward going diagonal lines in figure~\ref{fig:table} are of
the same order for $\as \ln^21/x\sim1$. On any given such diagonal,
the number of terms is always finite, due to the fact that the natural
hierarchy is single logarithmic. The leading terms in this
regime, discussed above in eq.~(\ref{eq:DLclass}), are highlighted by
the upper (upward-going diagonal) ellipse.  The lower ellipse contains
terms suppressed insofar as they have one less power of $\ln 1/x$,
\begin{equation}
  \label{eq:NDLclass}
  A_{k,2k-6}\, \asb^k \ln^{2k-6} \frac1x\,,\qquad\quad (3\le k\le 5)\,.
\end{equation}
Equivalently, since we are interested in the region where $\ln 1/x \sim
1/\sqrt{\as}$, these terms are suppressed by a power of $\sqrt{\as}$.
Adding the terms of the lower ellipse to eq.~(\ref{eq:firstterms}) one
obtains
\begin{equation}
  \label{eq:PggwithA42}
  x P_{gg}(x) = \mathrm{const.} + A_{31}\, \asb^3 \ln
  \frac1x + A_{43}\, \asb^4 \ln^3\frac1x + A_{42}\,\asb^4
  \ln^2\frac1x  + \order{\asb^k\ln^{2k-7}\frac1x}\,,
\end{equation}
where we have exploited the fact that $A_{54}=0$ and that the $A_{30}$
contribution can be absorbed into the constant piece. Solving for the
minimum of eq.~(\ref{eq:PggwithA42}) gives 
\begin{align}
  \ln \frac1{x_{\min}} 
  \label{eq:xminSLN}
  &= 
  \sqrt{-\frac{A_{31}}{3A_{43}\asb} + \frac{A_{42}^2}{9A_{43}^2}} -
  \frac{A_{42}}{3A_{43}}
  \\
  \label{eq:xminNL}
  &= \sqrt{-\frac{A_{31}}{3A_{43}}\frac1{\asb}}
  - \frac{A_{42}}{3A_{43}} + \order{\sqrt{\as}}
  \simeq \frac{1.156}{\sqrt{\asb}} + 6.947 + \order{\sqrt{\asb}}
  \,,
\end{align}
where the numerical values have been given for $\nf=4$.
We see that the effect of the subleading $A_{42}$ term is to shift
$\ln 1/x_{\min}$ by a (rather large) constant. 

As well as considering the position of the dip, it is interesting to
study also its depth, $d$. Substituting $\ln 1/x \sim \asb^{-1/2}$
into eq.~(\ref{eq:firstterms}) one immediately sees that the dip's
depth is of order $\asb^{5/2}$. Including the subleading terms (lower
ellipse of figure~\ref{fig:table}) gives the following result
\begin{equation}
  \label{eq:Pggdepth}
  -d =  \frac{2A_{31}}{9}
  \sqrt{\frac{-3A_{31}}{A_{43}}}\asb^{5/2} - 
  \frac13 
  \frac{A_{31}A_{42}}{A_{43}} \asb^3 + \order{\asb^{7/2}}
  \simeq -1.237\asb^{5/2} - 11.15\asb^3 + \order{\asb^{7/2}}\,.
\end{equation}
The depth has been defined with respect to the $x=1$ limit of
eq.~(\ref{eq:PggwithA42}), which includes the usual $\asb$ constant
term, but also $A_{20} \asb^2$ term and the unknown N\NLLx term
$A_{30} \asb^3$. The full $P_{gg}$ splitting function has of course a
$1/(1-x)_+$ divergence so its $x=1$ value can not actually be used as
a reference point for defining the depth. So one may choose to define
it alternatively with respect to the value of the $x\to0$ LO splitting
function, $A_{10}\asb$. This introduces extra terms $A_{20} \asb^2 +
A_{30}\asb^3$ in the expression, eq.~(\ref{eq:Pggdepth}), for $-d$.

\begin{figure}[tb]
  \centering
  \includegraphics[width=0.472\textwidth]{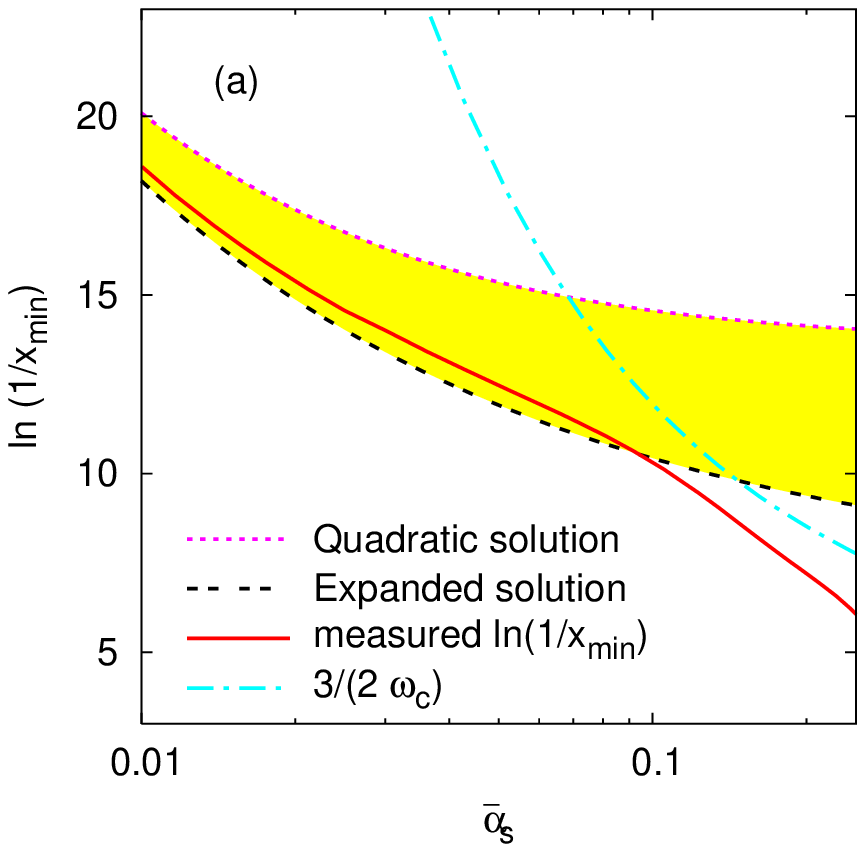}
  \hfill
  \includegraphics[width=0.49\textwidth]{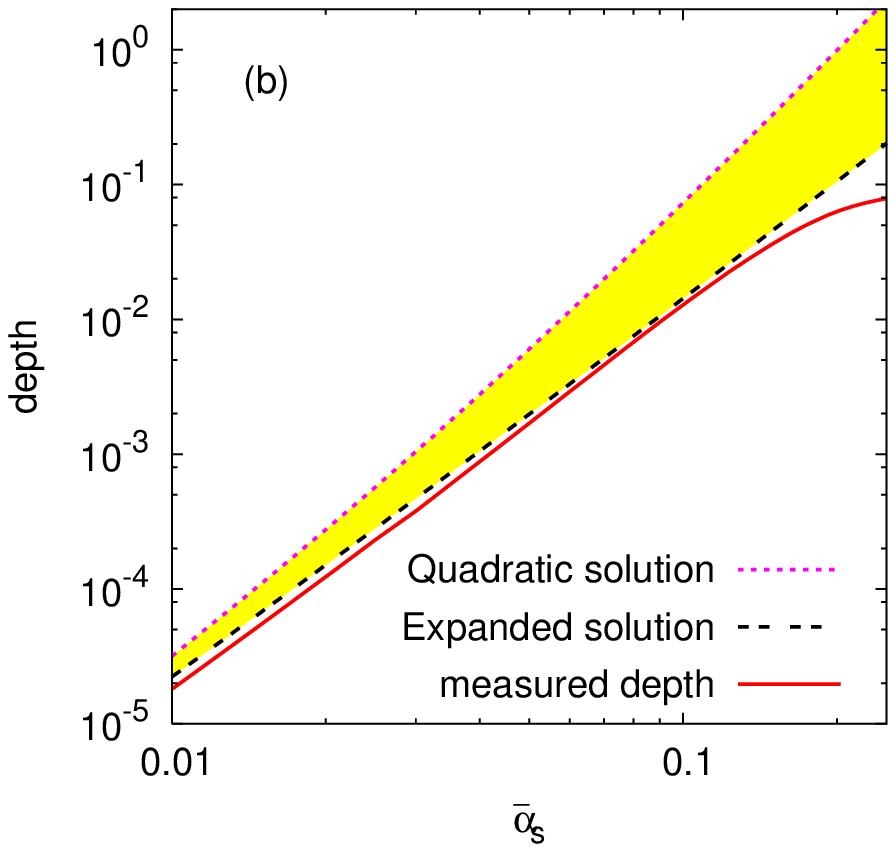}
  \caption{Properties of the dip in the \NLLB model of~\cite{CCSSkernel}
    compared to our analytical predictions.  See
    text for details.}
  \label{fig:dipprop}
\end{figure}

The dip position and depth, as a function of $\asb$, are shown
respectively in figures~\ref{fig:dipprop}a and \ref{fig:dipprop}b. In
each case the solid line represents the dip properties as `measured'
from the \NLLB scheme\footnote{We note that since the \NLLB scheme
  accounts only partially for the $\nf$ dependence (that associated
  with running of the coupling), the resulting \NLLx $A_{n,n-2}$
  coefficients differ slightly from those shown in
  eq.~(\ref{eq:Bcoeffs}), with $A_{20}$, $A_{31}$ and $A_{42}$
  corresponding to the $\nf=0$ results of eq.~(\ref{eq:Bcoeffs}) (in
  $A_{42}$ the $\nf$-part in the $b$-dependent term is retained, hence
  the coefficient of $\nf$ in $A_{42}$ is $0.334$). The reason for the
  only partial inclusion of the $\nf$ dependence is that the \NLLB
  scheme is based on a single-channel, purely gluonic approach,
  whereas full account of $\nf$ dependence would require a
  two-channel, quark-gluon formulation.}
of~\cite{CCSSkernel}, which was shown also in figure~\ref{fig:pgg}.
The shaded band represents the spread of the predictions based on
eqs.~(\ref{eq:PggwithA42})--(\ref{eq:Pggdepth}).  The upper edge of
the bands, labelled `Quadratic solution' corresponds to the use of
eq.~(\ref{eq:xminSLN}) and its direct substitution into
eq.~(\ref{eq:PggwithA42}); the lower edge corresponds to
eqs.~(\ref{eq:xminNL}) and (\ref{eq:Pggdepth}). For small values of
$\as$, there is rather good agreement between the expanded forms of
our predictions and the dip properties as measured from the full
resummation: the dip position is within the uncertainty band,
typically close to the expanded solution; the depth is just outside
the uncertainty band (again closer to the expanded solution), though
this may be because we have measured the depth with respect to the
$A_{10} \as$ reference level and have not included the resulting
additional unknown N\NLLx $A_{30}$ contribution to the depth.
Instead, including the $A_{30}$ as it appears in the \NLLB model,
lowers the band so that it overlaps with the measured depth. Leaving
aside these details, for both the position and depth of the dip, the
scaling with $\as$ is clearly reproduced, providing strong evidence
that the dip truly is a consequence of the low-order behaviour of the
perturbation series.

We note though that the spread of predictions, based on
eqs.~(\ref{eq:xminSLN}) and (\ref{eq:xminNL}), is quite significant.
This is essentially due to the large value of the $A_{42}$
coefficient, which means that the series in $\sqrt{\as}$ in
eqs.~(\ref{eq:xminNL}) and (\ref{eq:Pggdepth}) is very poorly
convergent --- the leading and subleading corrections are of the same
order when $\as \sim 0.01$--$0.02$.

In practice our low-order arguments seem to extend somewhat further,
providing a reasonable description of the dip, within the large
uncertainties, up to $\as \sim 0.05 - 0.1$. However beyond this point
the prediction fails quite dramatically, with the height of the
predicted dip minimum becoming for example negative ($A_{10}\asb - d
<0$), in contradiction with the full resummed results. Furthermore
there is a clear change in the $\as$ dependence for both the measured
position and depth of the dip.  This suggests that for $\as \gtrsim
0.05$ the dip description can no-longer be founded on low-order
perturbation theory alone.

\section{Resummation and cut-representation argument}
\label{sec:resummation}

On the other hand we know that when $\as$ is moderate and $\ln 1/x$ is
sizeable we enter the usual regime of resummation of terms $(\as \ln
1/x)^n$~\cite{BFKL}, together with its subleading
corrections~\cite{NLLFL,CC}. Though the strict \LLx, \NLLx hierarchy
is
ill-behaved, the inclusion of renormalisation group effects tends to
stabilise this hierarchy (e.g.~\cite{CCSSkernel,ABF2003}). As a
result one obtains the usual, expected behaviour of a splitting
function that increases as a power of $x$ at small $x$.

A simple estimate of the $x$ value for which this increase occurs can
be obtained in the approximation of a frozen coupling using the
quadratic expansion of the effective BFKL characteristic function
\begin{equation}
  \label{eq:quadchi}
  \asb \chi_{\eff}(\gamma,\asb) = \oms(\asb)(1 + D(\asb)(\gamma -
  \gamma_m)^2)\,, 
\end{equation}
where $\oms(\as)$ is the value of $\as\chi$ at its minimum, $\gamma =
\gamma_m$, and $D(\as)$ is related to the second derivative of $\chi$
(see figures~1 and 3 of~\cite{CCSSkernel}).

This leads to the well-known square-root branch-point for  the
anomalous dimension,
\begin{equation}
  \label{eq:branchpoint}
  \gamma = \gamma_m + \sqrt{\frac{\om - \oms}{D\oms}}\,,
\end{equation}
and to the representation 
\begin{equation}
  \label{eq:branchrep}
  xP_{gg}(x) \simeq \int^{\oms(\asb)} \frac{d\om}{\pi} \sqrt{\frac{\oms
      - \om}{D\oms}} x^{-\om} \simeq \frac{x^{-\oms}}{
    2\sqrt{\pi \oms D}\, \ln^{3/2} 1/x}\,,
\end{equation}
for the splitting function. 

It is amusing to note that the above estimate shows a dip at
\begin{equation}
  \label{eq:omsdip}
  \oms(\asb) \ln \frac1x = \frac{3}{2}\,,
\end{equation}
due to the logarithmic prefactor. Of course the actual cut structure
of the anomalous dimension is much more complicated, showing a variety
of subleading branch cuts, generally at complex $\om$
values~\cite{EHW,BLUMVOGT}, which are needed to match the small-$x$
representation (\ref{eq:branchrep}) to perturbation theory for small
$\as \ln 1/x$.  For this reason the dip structure (\ref{eq:omsdip}),
based on the moderate-$x$ behaviour of (\ref{eq:branchrep}) is not
always to be taken seriously.\footnote{For example, as we have
  mentioned earlier, the fixed-coupling \LLx splitting function has no
  dip at all. It is interesting also to note that LL evolution with
  (subleading) 
  running coupling corrections does have a dip \cite{THORNE,CCS00} --- its
  small-$\as$ properties are different from those of the full \NLLx
  dip, because it is due to an interplay between terms $\as^4\ln^n1/x$
  ($1\le n\le3$) and so, in the limit of small $\as$ occurs for $\ln
  1/x$ of order $1$. 
  A related running-coupling LLx dip has been obtained in
  \cite{ABF2001,ABF2003}, though the different scale of the running
  coupling and the use of the Airy extrapolation mean that it has
  different formal small-$\asb$ properties from \cite{THORNE,CCS00}.}

However, in our resummed calculation, the existence of the dip relies
on the negative $\ln 1/x$-slope of the splitting function which is
pretty well represented by the $\sqrt{\as}$-expansion, as noticed
before.  Furthermore, for $\oms \ln 1/x \gtrsim 3/2$,
eq.~(\ref{eq:branchrep}) is a reasonable representation of the
splitting function and in cases --- as ours --- in which there is a
dip, we can take eq.~(\ref{eq:omsdip}) as an
\emph{upper bound} on its position.

In the running-coupling case it is to be kept in mind that the cut is
actually broken up into a series of poles, the leading one being at a
position $\omc(\as)$ which lies somewhat below $\oms(\as)$ (see figure
18 of~\cite{CCSSkernel}) because of running coupling effects.
Nevertheless, as long as $x$ is not too
small, the inverse Mellin transform (\ref{eq:branchrep}) does not
resolve the difference between a cut and series of poles.

Therefore, by joining the $\sqrt{\as}$-expansion with the
cut-representation arguments, we are led to believe that the
perturbative and resummed regions can be matched by the inequality
\begin{equation}
  \label{eq:ineq}
  \ln \frac1{x_{\min}} \simeq \frac{c_1}{\sqrt{\asb}} + c_2 \;\lesssim\;
  \frac{3}{2\omc(\asb)}\,,
\end{equation}
where $c_1$ and $c_2$ are provided by eq.~(\ref{eq:xminNL}), and we
have replaced $\oms$ with $\omc \lesssim \oms$.
Since the right-hand expression goes
as $1/\asb$, this equation provides a transition point in $\asb$, below
which one should use the perturbative (double-logarithmic)
representation described before, and above which one should use the
full resummed behaviour.

This is confirmed by the moderate $\asb$ region of
figure~\ref{fig:dipprop}a, where one sees a clear bend in the
behaviour of $\ln 1/x_{\min}$ when $3/2\omc$
becomes of the same order 
as the perturbative representation, eq.~(\ref{eq:xminNL}), with the
measured $\ln 1/x_{\min}$ remaining consistently below $3/2\omc$.

\section{Conclusions}
\label{sec:conclusion}

The arguments provided in this letter go some way towards explaining
the features of the dip for a range of $\as$ values, both in terms of
a perturbative series in powers of $\sqrt{\as}$ for small $\as$, and
in terms of a resummed upper bound of the dip position, $\sim
3/(2\omc)$, for moderate values of $\as$.

It is the moderate-$\as$ region that remains the least well
understood, the matching of the small-$x$ increase to the initial
decrease being a quite non-trivial problem.  For example the simple
resummed treatment given above is subject to additional
running-coupling effects (e.g.\ difference between $\omc$ and $\oms$)
which may contribute further displacement of the dip and which have
not been considered here. Nevertheless, the arguments given so far
show that the dip does exist, as a moderately small-$x$ phenomenon,
under the simple condition that the small-$x$ part of $xP_{gg}(x)$ has
initially a negative $\ln 1/x$ slope, as is the case starting at NNLO.

An important phenomenological point that remains to be made concerns
the validity of fixed-order expansions of the splitting functions.
From our analysis of the dip properties, it is clear that for $\as
\gtrsim 0.05$ one starts to see a breakdown of the perturbative
expansion.  Despite this fact one notes a remarkable property of
figure~\ref{fig:pgg}, namely that the pure NNLO expansion of the
splitting function coincides rather well with the resummed result up
to the position of the dip minimum --- considerably beyond the point
 in $x$ where one would have naively expected the $\as^4$ DGLAP terms to
completely change the behaviour of the splitting function. This holds
for a wide range of $\as$.

We cannot claim to have fully understood
this observation, however it does suggest that it may in general be
safe to use the fixed order, NNLO, $P_{gg}(x)$ splitting function down
to $x$ values corresponding to the dip position, and only beyond this
point will small-$x$ resummation be strictly necessary. Thus one can
use the `measured' dip position, the solid curve of
figure~\ref{fig:dipprop}a, as an estimate of the limit of validity
of the NNLO expansion at small $x$.  Considering this in the context
of the available $F_2$ data, one sees that the limit cuts through the
HERA kinematical range, suggesting that while much of the data will be
in the region that is `safe' for an NNLO analysis, there is also a
substantial region at lower $x$ and $Q^2$ in which resummation will be
needed.

\section*{Acknowledgments}

We wish to thank Guido Altarelli and Stefano Forte for several
stimulating discussions on the subject of small-$x$ splitting
functions.


\end{document}